\documentclass[10.5pt, conference]{IEEEtran}
\IEEEoverridecommandlockouts
\usepackage{cite}
\usepackage{amsmath,amssymb,amsfonts}
\usepackage{algorithmic}
\usepackage{graphicx}
\usepackage{textcomp}
\usepackage[T1]{fontenc}
\usepackage[utf8]{inputenc}
\usepackage{xcolor}
\def\BibTeX{{\rm B\kern-.05em{\sc i\kern-.025em b}\kern-.08em
    T\kern-.1667em\lower.7ex\hbox{E}\kern-.125emX}}
\begin{document}
%%%%%%%%%%%%%%%%%%%%%%%%%%%%%%%%%%%%%%%%%%%%%%%%%%%%%%%%%%%%%%%%%%%%%%%%%%%%%%%
%%%%%%%%%%%%%%%%%%%%%%%%%%%% TITLE %%%%%%%%%%%%%%%%%%%%%%%%%%%%%%%%%%%%%%%%%%%%
\title{On Quantum-Assisted LDPC Decoding Augmented with Classical Post-Processing\\
%{\footnotesize \textsuperscript{*}Note: Sub-titles are not captured in Xplore and should not be used}

}

%%%%%%%%%%%%%%%%%%%%%%%%%%%%%%%%%%%%%%%%%%%%%%%%%%%%%%%%%%%%%%%%%%%%%%%%%%%%%%%
%%%%%%%%%%%%%%%%%%%%%%%%%%%% AUTHORS %%%%%%%%%%%%%%%%%%%%%%%%%%%%%%%%%%%%%%%%%%

\author{\IEEEauthorblockN{Aditya Das Sarma$^1$, Utso Majumder$^1$, Vishnu Vaidya$^2$, M Girish Chandra$^3$, \\ A Anil Kumar$^3$, Sayantan Pramanik$^2$}
			  \IEEEauthorblockA{$^1$Jadavpur University, $^2$TCS Incubation, $^3$TCS Research}\\ 
			  
		  	  aditya.41200@hotmail.com, utsomajumder@gmail.com, vaidya.vishnu@tcs.com, \\ m.gchandra@tcs.com, achannaanil.kumar@tcs.com, sayantan.pramanik@tcs.com
	  	  }

\maketitle
%%%%%%%%%%%%%%%%%%%%%%%%%%%%%%%%%%%%%%%%%%%%%%%%%%%%%%%%%%%%%%%%%%%%%%%%%%%%%%%
%%%%%%%%%%%%%%%%%%%%%%%%%%%% ABSTRACT %%%%%%%%%%%%%%%%%%%%%%%%%%%%%%%%%%%%%%%%%
\begin{abstract}
Utilizing present and futuristic Quantum Computers to solve difficult problems in different domains has become one of the main endeavors at this moment. Of course, in arriving at the requisite solution both quantum and classical computers work in conjunction. With the continued popularity of Low Density Parity Check (LDPC) codes and hence their decoding, this paper looks into the latter as a Quadratic Unconstrained Binary Optimization (QUBO) and utilized D-Wave 2000Q Quantum Annealer to solve it. The outputs from the Annealer are classically post-processed using simple minimum distance decoding to further improve the performance. We evaluated and compared this implementation against the decoding performance obtained using Simulated Annealing (SA) and belief propagation (BP) decoding with classical computers. The results show that implementations of annealing (both simulated and quantum) are superior to BP decoding and suggest that the advantage becomes more prominent as block lengths increase. Reduced Bit Error Rate (BER) and Frame Error Rate (FER) are observed for simulated annealing and quantum annealing, at useful SNR range - a trend that persists for various codeword lengths.
%Correction: Add Classical post-processing, at useful SNR range, we can obtain better BER, FER as compared to classical methods 
\end{abstract}
%%%%%%%%%%%%%%%%%%%%%%%%%%%%%%%%%%%%%%%%%%%%%%%%%%%%%%%%%%%%%%%%%%%%%%%%%%%%%%%
%%%%%%%%%%%%%%%%%%%%%%%%%%%% KEYWORDS %%%%%%%%%%%%%%%%%%%%%%%%%%%%%%%%%%%%%%%%%
%%%%%%%%%%%%%%%%%%%%%%%%%%%%%%%%%%%%%%%%%%%%%%%%%%%%%%%%%%%%%%%%%%%%%%%%%%%%%%%
%%%%%%%%%%%%%%%%%%%%%%%%%%%% INTRODUCTION %%%%%%%%%%%%%%%%%%%%%%%%%%%%%%%%%%%%%
\section{Introduction}
% This document is a model and instructions for \LaTeX.
% Please observe the conference page limits. 

Low Density Parity Check (LDPC) codes are linear
block codes originally proposed in the 1960s by Gallager
in his seminal doctoral work. The name reflects the fact that the 
parity check matrix used in LDPC coding is sparse with low density of 1s in the matrix. The performance of the LDPC codes approach theoretically described capacity
limits, and therefore are very powerful. LDPC codes
have established themselves as appropriate candidates for
wireless systems based on multi-antenna multi-carrier
transmission. Suitably designed LDPC codes are also
proven to be excellent candidates for Hybrid Automatic
Repeat Request (HARQ) schemes. The success and the
consequent popularity of the LDPC codes over the years
has resulted in support and proposals for its utilization
in various applications and standards. Some examples are DVB-S2 (2nd
Generation Digital Video Broadcasting via Satellite), 5G
New Radio (NR) access technology standards, recent
revisions of the 802.11Wi-Fi protocol family and various
storage applications. Practically utilizable codes should
constitute certain favourable properties, especially low
encoding and decoding complexities, good waterfall
regions, low error floors and flexibility in the context
of getting different rates and frame lengths. There
are various code designs available, starting from the
pseudo-random constructions to sophisticated algebraic
and graph-based techniques. See \cite{johnson_2009}, \cite{5397612} and \cite{bae_abotabl_lin_song_lee_2019} and
some of the original references therein for more details.

Good performance of LDPC codes can be achieved
with a proper choice of code and decoding algorithm.
Belief Propagation algorithms, like the Sum-Product
algorithm are widely used in classical LDPC decoding.
The Sum-Product algorithm can be viewed as a message
passing algorithm operating on the Tanner graph,
which is a bipartite graph representing the parity check
matrix, and consisting of variable nodes and check (or
constraint) nodes. Each iteration of the algorithm can
be divided into two halves. In the first half, message
is passed from each check node to all adjacent variable
nodes and in the second, from each variable node to
its adjacent check nodes. The decoding performance
is achieved through multiple iterations of the message
passing along the edges of the graph, until some stopping
criterion is reached. In the direction of reducing
the complexity of the (regular) Sum-Product algorithm,
many variants of it have been proposed in the literature,
one example being, min-sum algorithm (see \cite{johnson_2009} and the references there in for details).

Currently, we are in an exciting period in Quantum
Technologies. With the intermediate-scale commercial
quantum computers becoming increasingly available,
Quantum Information Processing is witnessing spectacular
developments (see \cite{pramanik2021quantumassisted}, \cite{10.1145/3341302.3342072}, \cite{Preskill2018quantumcomputingin} and the relevant references
there in). Before quantum processors become scalable,
capable of error correction and universality \cite{10.1145/3341302.3342072}, the
current and near-term devices, referred to as the Noisy
Intermediate-Scale Quantum (NISQ) \cite{Preskill2018quantumcomputingin} devices are
getting explored for solving certain hard problems to
achieve significant speedups over the best known classical
algorithms. Promising results are already reported
for solutions in the areas like, optimization, machine learning and chemistry. Apart from speedup considerations,
quantum mechanical properties of superposition,
entanglement and interference are being explored for
solving problems differently with possible performance
improvements. In the NISQ era, the hybrid quantum-classical
processing has established itself as an essential
combination, and this “cooperation” will continue for a
long time.

Considering the hardness and complexity of the some
of the important problems in the current and emerging
Communication Systems, research efforts have been
under way to explore Quantum Computing paradigms
to solve them. Some references in this direction are
\cite{10.1145/3341302.3342072}, \cite{10.1145/3372224.3419207}, \cite{10.1145/3447993.3448619}, \cite{9520378}, \cite{app10207116} and \cite{9569339}, among many others.
Needless to say, due to the present requirements of
Quantum Computers (QCs), like dilution refrigerators
to maintain superconducting cooling, the usage of QCs
are targeted to the Centralized Data Centers (Radio
Access Networks), see for example, \cite{10.1145/3341302.3342072} and \cite{9520378}. In this
paper, similar to some of the references mentioned in
this paragraph, we would be considering the baseband
processing, in particular the LDPC decoding (in fact,
we use \cite{10.1145/3372224.3419207} as the starting point). The relevance of
LDPC codes in modern wireless networks can be seen
in the search for computationally efficient decoders and
their ASIC/FPGA implementations in \cite{10.1145/3372224.3419207}. As a futuristic
notion, it is also useful to see how Quantum Processing
Unit (QPU) enhanced (or accelerated) processing together
with the classical computation can be worked out
to carry out some of the complex and computationally
heavy processing at the data center.

It has been well established for the last few years that
QCs can “naturally” solve the discrete combinatorial optimization
problems. Many of these problems fall under
the unifying model of Quadratic Unconstrained Binary
Optimization (QUBO) (see \cite{Kochenberger2006AUF}). One of the approaches
to finding the solution to a QUBO formulation is to
construct a physical system, typically a set of interacting
spin particles (two-state particles) whose lowest energy
state encodes the solution to the problem, so that solving
the problem is equivalent to finding the ground state of
the system. Two main approaches have been identified
to find the ground state of interacting spin systems
(quantum optimization) on NISQ processors \cite{pramanik2021quantumassisted}, \cite{10.1145/3341302.3342072}:
Quantum Annealing (QA) and Quantum Approximate
Optimization Algorithms (QAOA) \cite{farhi2015quantum}. QA is an approximate or non-ideal implementation of Adiabatic Quantum Computing, which is an analog quantum computation. QA has been developed theoretically
in the early nineties but realized experimentally in a
programmable device by D-Wave Systems, nearly two decades later.
A digitized version of Quantum Adiabatic Computing leads to QAOA, a gate-circuit based quantum computing.

In this paper, we have taken up the study of LDPC
Decoding using Quantum Annealers similar to \cite{10.1145/3372224.3419207}. 
% The following are novel, albeit, incremental contributions. Keeping in mind the tandem working of Quantum and Classical computers, we have attempted to exploit the inherent randomness of the QCs and the outputs or the results of the “shots” are subjected to classical postprocessing to arrive at better inference (in particular, better decoding) performance. In fact, different outputs emerging from the shots are seen as a kind of “diversity”, which to the best of our knowledge are not interpreted this way in the existing literature. Further, these are postprocessed using simple minimum distance computation to the received codeword vector to arrive at the final decoding. Preliminary results with length 32 and 96 rate half LDPC codes (\cite{ccdb}) demonstrate the improved performance of the quantum-enhanced decoders, even with these short lengths, over conventional Sum-Product Algorithm. The paper also captures certain remarks about these new kind of formulations and is organized as follows: In Section II, we capture aspects related to classical Sum-Product and Min-Sum algorithms, QUBO and Annealing (both Simulated and Quantum). Section III provides the details about the Proposed augmented method; results and the discussions are covered in Section IV.
But, the following novelties are brought in. Keeping in mind the tandem working of Quantum and Classical computers, we have attempted to exploit the inherent randomness of the QCs and the outputs or the results of the “shots” are subjected to classical postprocessing to arrive at better inference (in particular, better decoding) performance. In this direction, instead of just picking up the minimum-energy solution as prevalent in the Quantum Computing literature, the different outputs are post-processed using simple minimum distance computation to the received codeword vector to arrive at the final decoding. This approach sets the direction to consider appropriate and more sophisticated post processing for Quantum-Enhanced baseband processing. We have taken this route to bring out a notion of diversity emerging from the shots. In fact, different outputs emerging from the shots are seen as a kind of “diversity”, which to the best of our knowledge are not interpreted this way in the existing literature. Preliminary results with length 32 and 96 rate half LDPC codes (\cite{ccdb}) demonstrate the improved performance of the quantum-enhanced decoders, even with these short lengths, over conventional Sum-Product Algorithm. The paper also spell out certain new remarks/observations about different formulations considered and is organized as follows: In Section II, we capture aspects related to classical Sum-Product and Min-Sum algorithms, QUBO and Annealing (both Simulated and Quantum). Section III provides the details about the Proposed augmented method; results and the discussions are covered in Section IV.

\section{Brief Elaboration on Quantum Annealer and QUBO} 
%Correction: Little elaboration on quantum annealer and QUBO
%%%%%%%%%%%%%%%%%%%%%%%%%%%%%%%%%%%%%%%%%%%%%%%%%%%%%%%%%%%%%%%%%%%%
\begin{figure*}[t!]                                                
    \begin{center}                                                 
        \includegraphics[width=1.99\columnwidth]{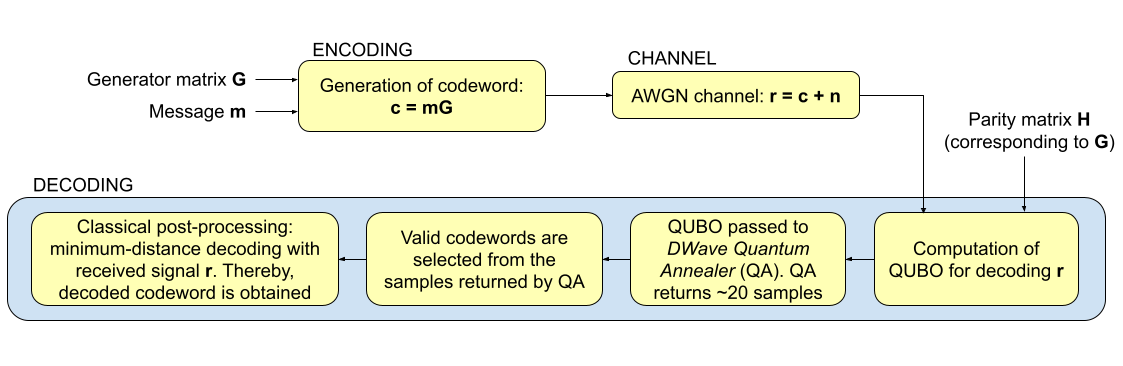}  
        \caption{Schematic of the approach}                    
        \label{fig:flowchart}                                
    \end{center}                                                   
\end{figure*}                                                       
%%%%%%%%%%%%%%%%%%%%%%%%%%%%%%%%%%%%%%%%%%%%%%%%%%%%%%%%%%%%%%%%%%%%%%%
% \subsection{Quantum Computing in Communication Networks}
% Over the past few years, quantum computing and its associated approaches have been employed in the field of signal decoding. In \cite{9366137},
% formulation of an ML problem using QUBO has been demonstrated, simulating the decoding on the D-Wave 2000Q annealer. They proposed two implementations with Ising model formulations, generated from the generator matrix and the parity-check matrix respectively, both of which were established o be performing better for small code length, over belief propagation (BP) decoding and brute-force ML decoding with classical computers. 
% Similarly, hard decision problems are being solved using quantum methods. An implementation of quantum annealing techniques through two different general mappings of planning problems to quadratic unconstrained binary optimization (QUBO) problems, and applying them to two parameterized families of planning problems, navigation-type and scheduling-type, can be seen in \cite{rieffel2015case}. In \cite{10.1145/3341302.3342072}, quantum annealing has been implemented to a large MIMO centralized radio access network design, to provide better performance in multiuser systems.

%Correction: Related works is Redundant 

\subsection{D-Wave Quantum Annealer}
{Quantum Annealing} (QA) is a metaheuristic for solving QUBO problems\cite{hen2016quantum}. The adiabatic theorem of quantum mechanics states that Quantum Annealing, in a closed system, will find the final ground state encoding the solution, provided the annealing time is appropriately large compared to the inverse of the energy gap in quantum ground state. However, this does not guarantee that QA will always perform better than classical optimization algorithms, as the relative success of QA depends on the suitability of the optimization landscape to obtain an quantum advantage. D-Wave provides access to their devices which implement Quantum Annealing on Quantum hardware, through its cloud access provision Leap. Here, we are not capturing information on D-Wave Annealers, since nice documentation/information is available in their website. Also see \cite{PhysRevE.58.5355} \cite{finnila1994quantum}.

\subsection{QUBO}

The concept of a QUBO formulation is fundamental to utilizing a Quantum Annealer to solve a given optimization problem.

Let $f:B^{n} \rightarrow R$ be a quadratic polynomial with $q_i \in B = \{0, 1\}$ for $1 \leq i \leq n$:
\begin{equation}
    f_{\bar{\alpha}}(x) = \sum_{i=1}^{n}\sum_{j=1}^{i} \alpha_{ij} q_i q_j
\end{equation}
The QUBO problem then consists of finding $q^{*}$ such that:
\begin{equation}
    q^{*} = \underset{q \in B^{n}}{\text{argmin}}  f_{\bar{\alpha}}{(q)}
\end{equation}
The QUBO form of (1) can be written, separating the linear and quadratic terms, and noting that $q_{i}^2 = q_{i}$, and setting $\alpha_i = \alpha_{ii}$, as:
\begin{equation}
    f_{\bar{\alpha}}(q) = \sum_{i=1}^{n}\alpha_{i}q_i + \sum_{i=1}^{n}\sum_{j=1}^{i}\alpha_{ij}q_{i}q_{j}
\end{equation}
$\alpha_i$ is called the the bias of the variable $q_i$, and $\alpha_{ij}$ is called the bias/coupling of the quadratic term $q_i q_j$.

Any optimization problem that we wish to solve with the QA, must first be formulated as a QUBO problem. We discuss the QUBO formulation of LDPC decoding in the next section.

\section{Proposed Approach}
The flowchart  given in Fig. 1 summarizes the proposed solution approach. In the following sub-sections requisite details are elaborated.

\subsection{Encoding}
\begin{itemize}
    \item To implement the LDPC encoding, we consider a valid parity matrix $\mathbf{H}$ and the corresponding generator matrix $\mathbf{G}$. 
    \item For a randomly generated message \textbf{m}, codeword \textbf{c} corresponding to \textbf{m} is obtained by multiplying \textbf{c} with the generator matrix \textbf{G}.
    \begin{equation}
        \textbf{mG} = \textbf{c}
    \end{equation}
    where the multiplication is mod-2.
    \item To simulate the effect of the channel on the transmission of the codeword, we add Additive White Gaussian Noise (AWGN) to the transmitted codeword, to obtain the received signal \textbf{r}:
    \begin{equation}
        \textbf{r} = \textbf{c} + \textbf{n}
    \end{equation}
    where $\textbf{n} \sim  \mathcal{N}_{N}(\mathbf{0}, \sigma \mathbf{I})$. We can adjust SNR by adjusting the variance $\sigma$.

\end{itemize}

\subsection{Decoding}
%--------------------------------------------
\begin{figure*}
    \begin{center}
        \includegraphics[width=.99\columnwidth]{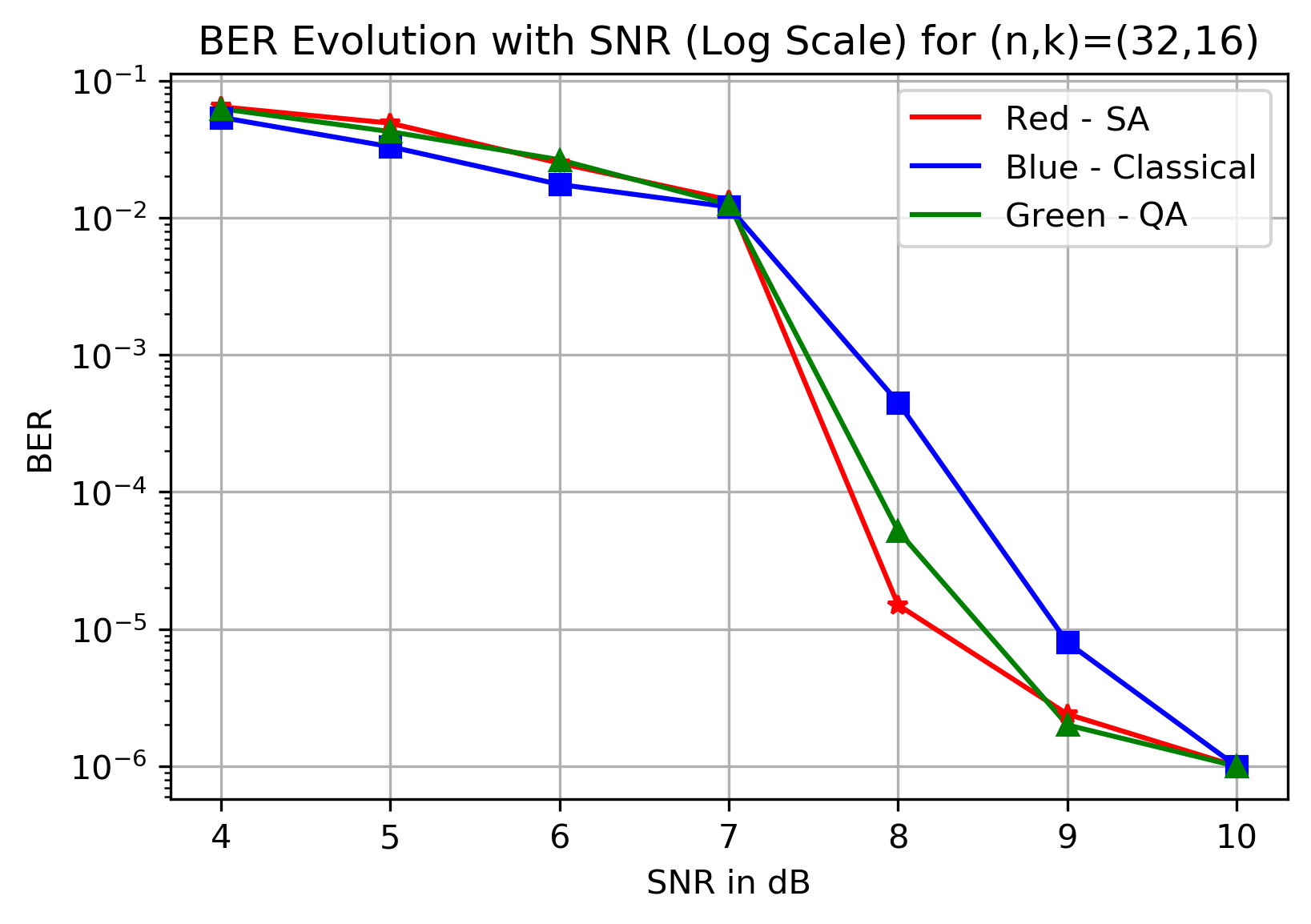}
        \includegraphics[width=.99\columnwidth]{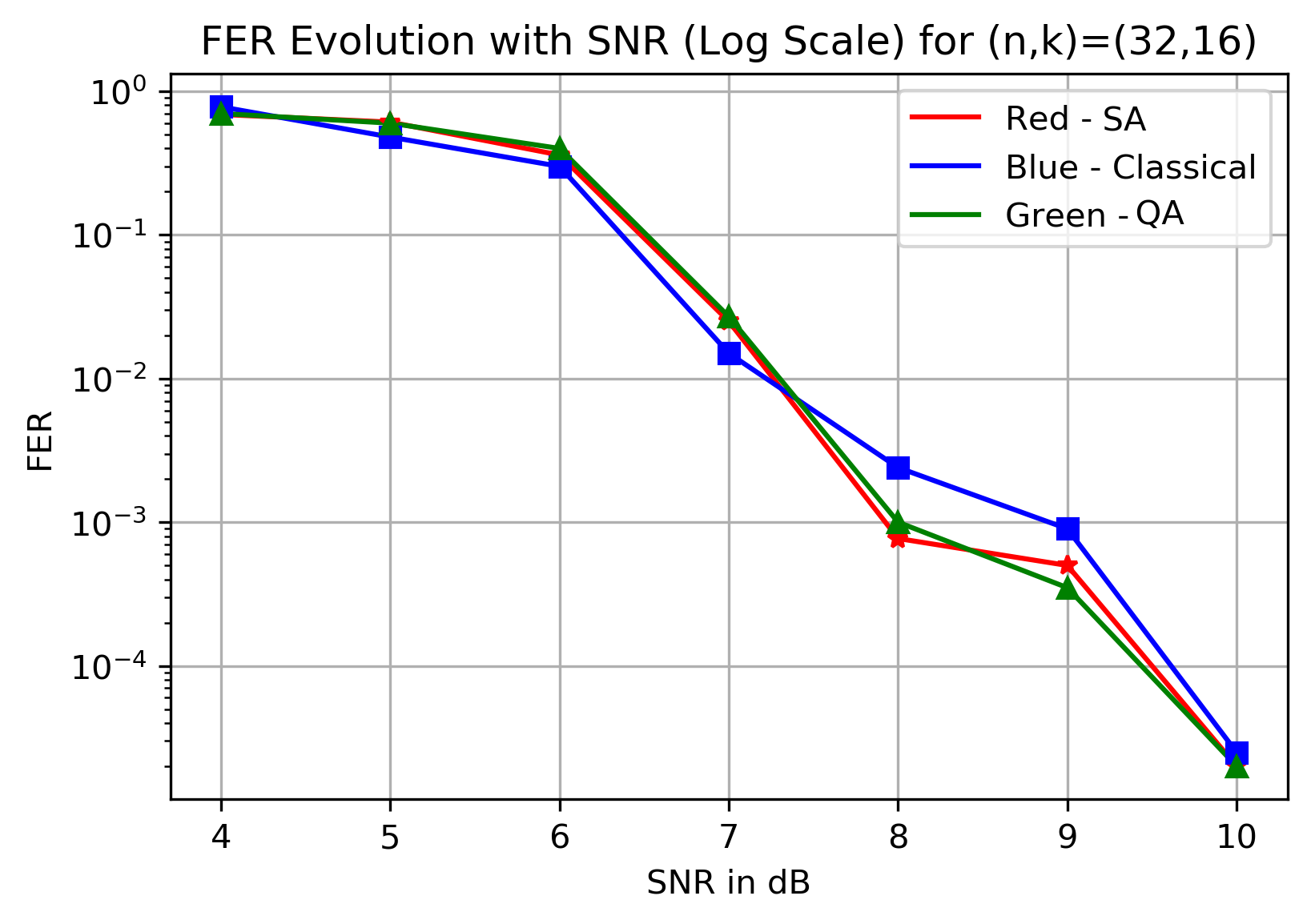}
        \includegraphics[width=.99\columnwidth]{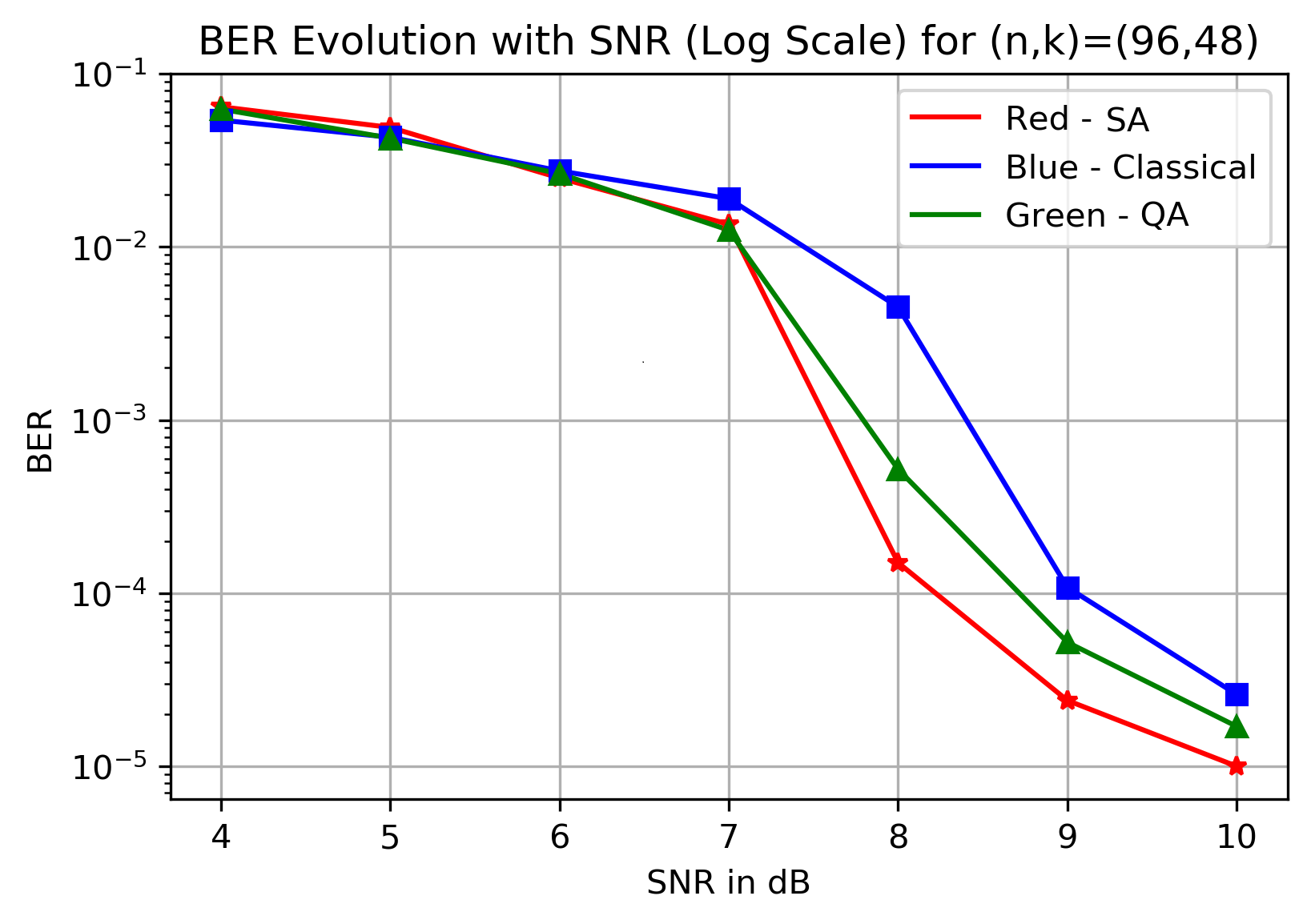}
        \includegraphics[width=.99\columnwidth]{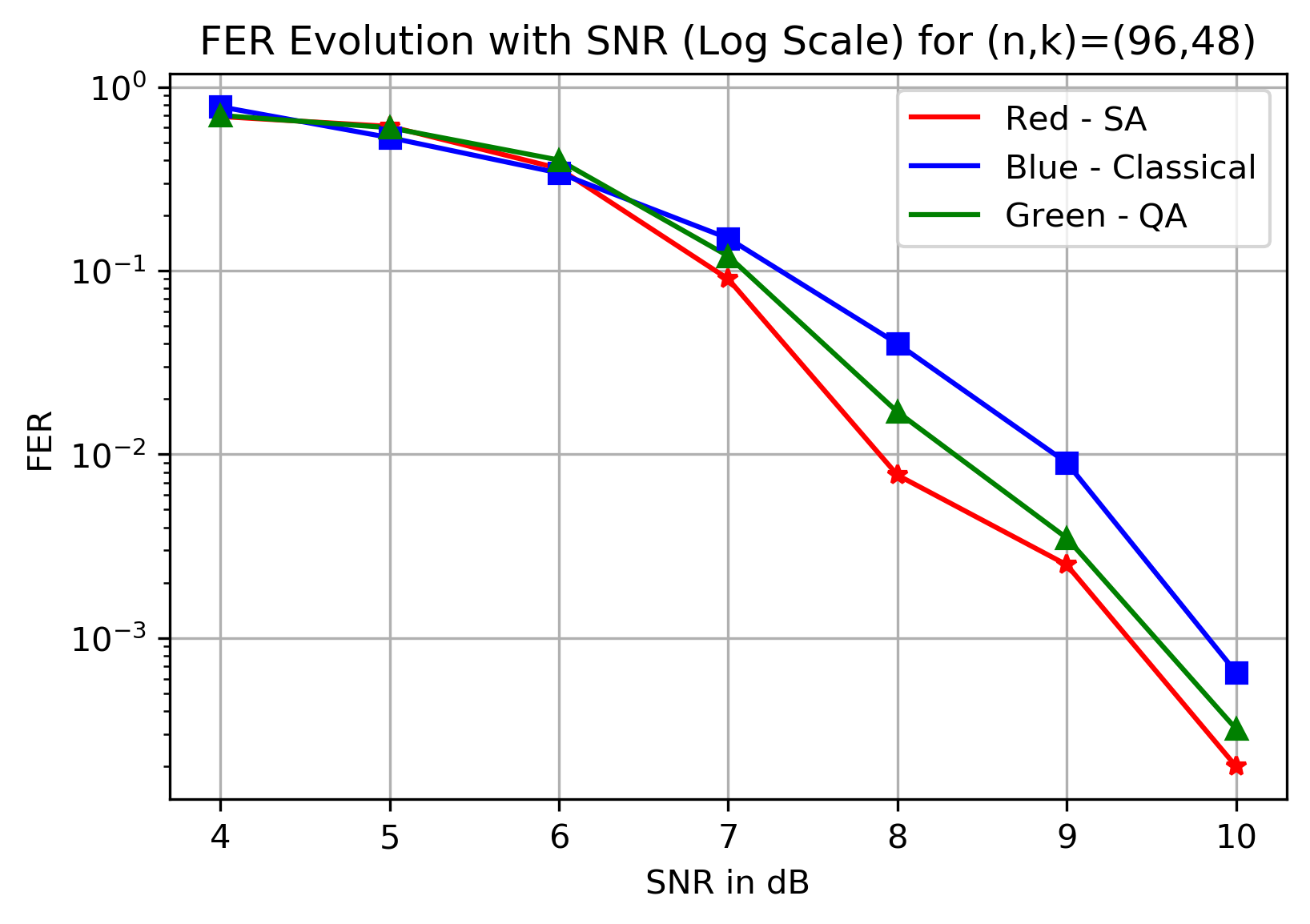}

    \hfill

    \end{center}
\caption{BER, FER vs SNR for different problems}
    \label{fig:overview}
\end{figure*}
%---------------------------------------------

\begin{itemize}

 \item To decode the received signal $\mathbf{r}$, we first put in place the corresponding QUBO formulation. The QUBO for $\mathbf{r}$ is composed of two parts: 
    \begin{enumerate}
        \item Distance Metric: 
    Let binary variable $q_i$ represent the $i^{th}$ bit of the decoded codeword. We compute the expectation of $q_i$ given the received symbol $r_i$, as $P(q_i = 1|r_i)$. For an AWGN Channel with Binary Phase Shift Keying (BPSK) Modulation, this quantity, as given in \cite{10.1145/3372224.3419207}, is:
    
    \begin{equation}
        Pr(q_i = 1|r_i)=\frac{1}{1+\exp{\frac{2r_i}{\sigma^2}}}
    \end{equation}
    %Correction: intrinsic LLR
    
    We expect that the transmitted codeword is "proximal" to the received signal. Therefore, to find the transmitted codeword, we seek to minimize the following Distance Metric $\delta$ that computes the proximity of a codeword to the received information:
    \begin{equation}
            \delta = \sum_{i=1}^{n} (q_i - Pr(q_i = 1|r_i))^2
        \end{equation}
    A minimum of (7) is an estimate of the transmitted codeword, computed with the quantities $Pr(q_i=1|r_i)$ alone.
    %Correction: The following is the objective function is to be minimised
    
    %Correction for BP: For proposed methods, the complicated optimization to arrive at the solution doesn't require iterations like in the classical methods.
    
        \item Constraint Satisfaction Metric: The LDPC constraints ensure that the modulo-2 sum at each check node $c_n$ is $0$. These equality constraints need to be incorporated into an objective function that can be minimized. We implement this with the following function. For each check node $c_i$ one can define LDPC satisfier function (see also [7]):
    %Correction: Mention Le and Lsat details
        \begin{equation}
            L_{sat}(c_i) = ((\Sigma_{\forall j:h_{ij}=1}q_j) - 2L_{e}(c_i))^2 
        \end{equation}
        Through minimization of the above function, we can force the sum at that check node to be even: that is, force the modulo-2 sum at that node to zero. $L_e(c_i)$ is implemented with additional ancillary qubits.
        Next enters the Constraint Satisfaction Metric $L$:
        \begin{equation}
            L = \sum_{i} L_{sat}(c_i)
        \end{equation}
        Minimizing $L$ would result in the satisfaction of the LDPC constraints at the check nodes.
    
    \end{enumerate}
    Finally, we combine the two components with Langrange weights $W_1$ and $W_2$, to compose the final QUBO. Minimizing the QUBO in general tends to minimize both the composite components. We can prioritize the minimization of one component over the other with a high choice for the Langrange weight for that component relative to the other. We have experimented with variations on $W_1$, keeping $W_2$ fixed at 1.0. The resulting QUBO is:
        \begin{equation}
            F = W_1 \delta + W_2 L
        \end{equation}

   \item The QUBO is then passed to the D-Wave annealer. Several samples are collected by running the annealer multiple times.
    \item Valid codewords (codewords that satisfy LDPC constraints) are filtered out from the samples and then minimum distance decoding is performed with the received signal to obtain the final decoded codeword.

\end{itemize}

As can be seen from the above description this QA-based framework doesn't require message passing iterations typically used to perform LDPC decoding with classical BP algorithms. Instead, a Quantum Annealer implemented on real Quantum hardware "naturally settles" to the optimal state for the QUBO, thereby performing the LDPC decoding.  
    
% \begin{figure}
%     \centering
%     \includegraphics[width=.99\columnwidth]{./Assets/cost2.png}
%     \caption{Caption}
%     \label{fig:cost}
% \end{figure}

\section{Experimental Results}
%Correction: Mention the site from which code matrices are taken
Decoding was performed on LDPC parity matrices of dimensions (32, 16) and (96, 48), using quantum and simulated annealing, and classical Belief Propagation algorithms (see \cite{ccdb}). 
%Correction: Mention the inherent diversity posed by quantum methods (due to the stochastic feature), which cannot be exhibited by classical, which end in same outcomes, for same given r
Quantum methods provide an inherent mode of diversity, due to its stochastic nature, giving different outputs for the same \textbf{r}, for different runs of the experiment. This advantage is not available for classical BP algorithms, which are deterministic in nature. In other words, for successive runs of the experiment, using the same \textbf{r} results in different outputs due to the inherent randomness in quantum information processing. On the other hand, it is trivial to observe that the same output, and not the "different copies" of information related to the transmitted codeword. Of course, this benefit is coming because of the use of Quantum Computers.

\subsection{Results for fixed SNR channel}
For this scenario, different SNRs are considered for experimentation. For each SNR, the BER and FER estimate is obtained with $10^{6}$ Monte Carlo iterations. The term "fixed" refers to the fact that the SNR remains the same for all these $10^{6}$ "transmissions". Based on the four plots in Fig. 2, the following observations are evident:
\begin{itemize}
    \item In the moderate SNR regime, Quantum Annealing (QA) and Simulated Annealing (SA) perform better than the classical BP.
    \item At lower SNRs, performance of QA and SA is close to the performance of classical BP. 
    \item However, a sharp drop is seen in BER, as well as in FER, around 7.5 to 8 dB range for both simulated and quantum annealing. When SNR reaches 10dB, the noise becomes small enough such that all the methods  achieves the similar BER and FER.
    
\end{itemize}
In the limited amount of studies we carried out using
two short codewords, Simulated Annealing performs
slightly better than Quantum Annealing. It is to be noted that the QA results are obtained from the actual D-Wave Annealer, and these realistic machines do have imperfections ("noisy behavior") at present.  Of course, as remarked earlier, both SA and QA performed better than classical BP.

\subsection{Results for time-varying SNR}

\begin{figure}
    \centering
    \includegraphics[width=.99\columnwidth]{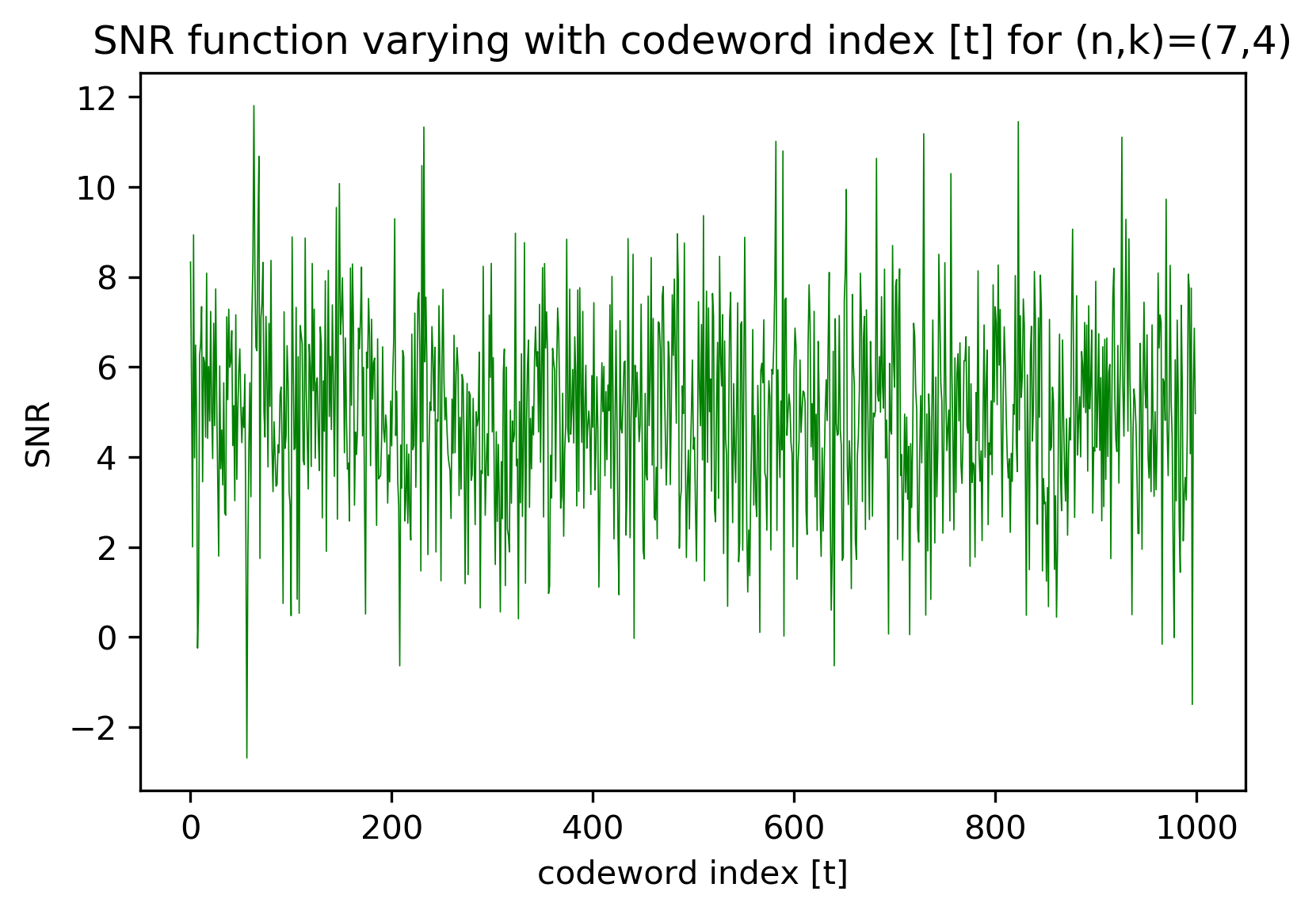}
    \caption{Variation of SNR with respect to time}
    \label{fig:SNR}
\end{figure}
%ZCorrection: Dont mention rayleigh. Replace with time-varying snr analysis
In order to simulate a time-varying SNR function and observe how the proposed approach performs in this case, the following procedure was undertaken and observations were recorded.
\begin{itemize}
    \item For each of the 1000 codewords transmitted, the SNR is varied. In our experimentation, the samples have been drawn from the normal distribution with $\mu$=5, $\sigma$=2. A realization of the SNR is depicted in Fig. 3
    \item It is again observed that simulated annealing has the highest fraction of correct codewords decoded, followed by quantum annealing and classical belief propagation, as given in Table \eqref{table:tab-gate2}.
\end{itemize}

In this paper, we have just considered a time-varying SNR to assess the performance of the proposed methodology. In the direction of considering the more realistic scenarios, we are in the process of implementing complex-baseband processing with the Rayleigh fading channel. The possible modifications to QUBO formulation for this case is also envisaged.

The results for both fixed and time-varying SNR demonstrated the correct functionality of the QUBO formulation of the LDPC decoding augmented with post-processing which exploits the special diversity mentioned. Experimentations with longer codewords may bring out the beneficial aspects of the proposed approach compared to classical counterparts. Elaborating further, it is expected that the Quantum Computers, including the Annealers will only improve in terms of number of qubits, quality of the qubits, the connectivity between them, etc. They can then not only accommodate larger-sized problems (for instance, longer code lengths of practical importance, etc), but also naturally solve them with better performance and speed compared to the fully classical counterparts (both Simulated Annealing and the variants of Sum-Product). Additionally, the right integration of classical and quantum computing systems may result in useful energy savings as well \cite{https://doi.org/10.48550/arxiv.2109.01465}.

\begin{table}[t]
\caption{Fraction of correct codewords for time-varying SNR (for $10^6$ Monte Carlo instances)}
\label{table:tab-gate2}
\centering
    \begin{tabular}{|c|c|}
        \hline
        \textbf{Methods} & \textbf{Fraction of correct codewords}\\
        \hline
        Classical Belief Propagation & $0.848$\\
        Simulated Annealing & $0.946$\\
        Quantum Annealing & $0.902$\\
        \hline
    \end{tabular}
\end{table}

\section{Conclusion \& Future Work}
Classical post-processing assisted quantum annealing is proposed for LDPC decoding which exploits the stochastic nature of quantum computers to arrive at improved solutions at SNRs of practical relevance, when compared with classical BP decoding. Unlike classical BP decoding, iterations are not required for this QA-based approach. The physical nature of the quantum computing allows natural settlement to the lowest energy state, which in most cases is the optimal solution.
%the complicated optimization to arrive at the solution doesn't require iterations like in the classical methods.
Post-processing was based on minimum distance decoding schemes, this can be extended to more refined methods. There is plenty of scope for expanding this work for different baseband processing techniques in future, including different channels such as Rayleigh fading channel. 

%Mention: Post-processing was based on minimum distance decoding schemes, this can be extended to more refined methods.Also mention different baseband processing in future.

%%%%%%%%%%%%%%%%%%%%%%%%%%%%%%%%%%%%%%%%%%%%%%%%%%%%%%%%%%%%%%%%%%%%%%%%%%%%%%%
%%%%%%%%%%%%%%%%%%%%%%%%%%%% REFERENCES %%%%%%%%%%%%%%%%%%%%%%%%%%%%%%%%%%%%%%%
\bibliographystyle{IEEEtran}
\bibliography{ref.bib}

\end{document}